\documentclass[aps,preprint]{revtex4}%
\usepackage{amsfonts}
\usepackage{amsmath}
\usepackage{amssymb}
\usepackage{graphicx}%
\begin{document}
\title{Andreev Reflection and Long-Range Proximity Effect in Pb/LaCaMnO Point Contacts}
\author{V. N. Krivoruchko, V. Yu. Tarenkov, A.I. D'yachenko, and V.N. Varyukhin}
\affiliation{Donetsk Physics \& Technology Institute NAS of Ukraine, Str. R. Luxemburg 72,
83114 Donetsk, Ukraine}
\date{\today}


\begin{abstract}
The Andreev reflection (AR) spectroscopy has been used to probe mutual
influence of superconducting pairing and ferromagnetic correlations in
contacts of low-temperature superconductor (S), Pb, and half-metallic
ferromagnetic, La$_{0.65}$Ca$_{0.35}$MnO$_{3}$ (LCMO). Two different AR
spectra have been observed. In the contacts without proximity effect, the AR
spectra reveal a behavior typical for superconductor/spin-polarized metal
contacts. In the contacts, which we distinguish as the proximity affected
ones, a few unusual effects have been detected. Namely, we have observed: (i)
a spectacular drop in the resistance of the contacts at the onset of the Pb
superconductivity; (ii) a typical for S/normal nonmagnetic metal contacts,
excess current and doubling of the normal-state conductance; and (iii) a much
larger than it should be for conventional superconductivity ratio of the
single-particle gap to the Pb superconducting transition temperature. The very
distinct AR characteristics of proximity affected contacts observed by us
suggest unconventional (triplet) superconducting pairing and long-range
proximity effect at S/LCMO interface.

\end{abstract}
\maketitle

\textit{Introduction}. - Electron transport in ferromagnetic (F)
metal/superconductor (S) heterocontacts has attracted considerable attention
recently. The key phenomenon that controls the behavior of such systems is
proximity effect [1-3]. Theory predicts that in the extreme case of a
completely spin-polarized material the conventional (singlet s-wave pairing)
proximity effect is absent [4]. Therefore one might expect that the influence
of the superconducting proximity effect on the transport properties of such
heterostructures should be negligibly small. A rapidly growing number of
experiments seem, however, to contradict these conclusions and show that a
novel type of proximity effect may be realized in S/F proximity coupled
structures. Kasai \textit{et al}. [5] were the first to our knowledge who
asserted this hypothesis based on the results of the investigation of
current-voltage characteristics for YBa$_{2}$Cu$_{3}$O$_{y}$/ magnetic
manganese oxide/YBa$_{2}$Cu$_{3}$O$_{y}$ junctions. For certain values of x,
the authors observed that supercurrent passes through a half-metallic
ferromagnetic (HMF) layers up to 200nm thickness. These results can not be
explained in paradigms of the conventional proximity effect and suggested a
novel long-range proximity effect concerned with magnetism of the barrier. An
unconventional mutual influence of superconductors and ferromagnetic
conductors in hybrid S/F nanostructures was also reported recently by a few
groups [6-11].

Despite the existing experimental evidences supporting unconventional
proximity effect in S/HMF structures, understanding of this phenomenon still
remains unclear and requires direct measurement of such intrinsic
characteristics of the superconducting state as the local quasiparticle
density of states (DOS), value of the superconducting gap, spin symmetry of
the pairs, etc. As is well known, such methods as tunnelling spectroscopy [12]
and Andreev reflection (AR) spectroscopy [13] are direct and sensitive tools
for these purposes. In particular, recently the crossed AR spectroscopy of a
metallic point contact has been discussed as a sensitive probe of
superconducting order parameter spin-symmetry [14,15].

In this report, first to our best knowledge, the AR spectroscopy has been
utilized to probe the unconventional superconducting pairing in contacts of
low-temperature superconductor and a half-metallic ferromagnet. If triplet
pairing exists in S/HMF heterostructures, the S/F point contacts should reveal
such typical for S/normal nonmagnetic metal (N)\ contacts distinct features as
excess current and doubling of the normal-state conductance, which have not
yet been demonstrated in experiment until now. Such material as Pb is used as
the superconducting electrode and La$_{0.65}$Ca$_{0.35}$MnO$_{3}$ (LCMO) as
the HMF. Two qualitatively different AR spectra have been detected. In the
contacts without visible proximity effect, the AR spectra reveal a behavior
typical for S/spin polarized metal contacts (see, e.g., Refs. [16-19]). In the
contacts with visible superconducting proximity effect, new phenomena have
been observed. Namely, a typical for S/N interface, excess current and
considerable increase (doubling) of the contact's conductance below the
superconducting critical temperature have been detected. Restored magnitude of
the single quasiparticle gap is anomalously larger than one can expect for
conventional BCS model $\Delta_{BCS}$(T=0)/T$_{C}$ ratio. Incorporating the
inherent magnetic nonhomogeneity of manganites, we speculate possible physical
mechanisms behind the observed unconventional features of the AR spectra. Our
results provide spectroscopic evidence for existence of triplet
superconducting pairing and long-range proximity effect at S/HMF boundary.

\textit{Experimental details.} -- We have studied point contacts with geometry
shown in Fig. 1. Textured LCMO plates were grown using standard ceramic
technique. In particular, ceramic powder plates sized 0.1$\times$1$\times$10
mm$^{3}$ were pressed (20 Kbar) and then subjected to annealing for 8h at
1250$^{0}$C. This leads to an increase of the average size values of
crystallites up to values about 5-10 $\mu$m. The resistivity was measured by
the standard four-probe method with the low frequency ac technique. A typical
low-temperature superconductor Pb has been used as a superconducting
electrode. Metallic contacts between LCMO plate and superconducting wire were
formed by pressing slide-squash up a needle-shaped Pb against the LCMO
surface. The contacts were made at room temperature and at liquid nitrogen, as
well, but the results did not depend upon the way of preparation. The
contacts' parameters were stable, offering a possibility to perform
measurements in wide temperature range. Note, that in comparison to
manganites/high-T$_{C}$ S structures in our Pb/LCMO contacts the so-called
hole-charge transfer effect [20] is not present.

Temperature dependence of the plates' resistance (see bottom insert in Fig. 2)
has a sharp maximum near T$_{Curie}$ $\approx$ 270K associated with the well
known metal-dielectric transition [21]. Below room-temperature the resistance
of the plates was $\sim$1$\Omega$. The low-field (H $\approx$ 100 Oe)
magnetoresistive effect [$\rho$(T,0) - $\rho$(T,H)]/$\rho$(T,0) at T = 77K was
only 0.3\% (not shown). This suggests that the contribution of inter granular
junctions to the total sample resistance is negligibly small. The transition
resistance of the current and potential contacts was R $\sim$10$^{-8}$
$\Omega\times$cm$^{2}$. The junctions resistance was much larger ($\sim$100
$\Omega$), so that the rescaling effects can be neglected.

The approximate value of the contact diameter \textit{d} is estimated by
employing the Wexler's formula [22]. We obtain \textit{d} $\sim$100\AA , that
is we deal with the so-called Sharvin contacts [23]. Taking into account small
contact's dimension we estimate the specific resistivity of the micro crystals
is of the order $\rho$ $\sim$10$^{-4}\Omega\times$cm, and for the mean free
path we obtain \textit{l} $\sim$100\AA . Since the mean free path \textit{l}
is an order of the contact diameter \textit{d}, this means that the transport
regime is into the intermediate region, etc., not a ballistic (\textit{l}
%TCIMACRO{\TEXTsymbol{>}}%
%BeginExpansion
$>$%
%EndExpansion%
%TCIMACRO{\TEXTsymbol{>} }%
%BeginExpansion
$>$
%EndExpansion
\textit{d}), and not a diffusive one (\textit{l}
%TCIMACRO{\TEXTsymbol{<}}%
%BeginExpansion
$<$%
%EndExpansion%
%TCIMACRO{\TEXTsymbol{<} }%
%BeginExpansion
$<$
%EndExpansion
\textit{d}). (Experimental details can be found in Ref. 24).

\textit{Results}. -- We recorded about a hundred of the AR spectra. Mostly, we
observed the well known spectrum that reflects the half-metallic properties of
LCMO. Representative normalized differential conductance G(V) = ($d$%
I/$d$V)(V/I) of the Pb/LCMO junction is shown in Fig. 3. As one can see, in
contrast to the conventional AR at S/N interface characterized by excess
current, here an excess voltage V$_{exc}$ is observed. The almost constant
V$_{exc}$ value is observed for
%TCIMACRO{\TEXTsymbol{\vert}}%
%BeginExpansion
$\vert$%
%EndExpansion
V%
%TCIMACRO{\TEXTsymbol{\vert} }%
%BeginExpansion
$\vert$
%EndExpansion
$\leq$ 20mV. This proves the suggestion that heating effects could be
neglected. Any visible features of superconducting proximity effect have not
been detected, and we will refer to these contacts as \textquotedblleft the
contacts without proximity effect\textquotedblright. For singlet pairing,
superconducting coherence length $\xi_{F}$ = ($\hbar$D$_{F}$/\textit{k}%
$_{B}\pi$T$_{Curie}$)$^{1/2}$ (in the dirty limit [1-3]) for LCMO is extremely
short $\sim$5$\div$7\AA \ (here D$_{F}$ denotes the diffusion constant in F).
Contribution of such a small region to the contact's resistance is less than 1\%.

The suppression of the AR in S/F point contacts has been observed
experimentally by several groups [16-19]. These results are qualitatively well
explained by de Jong and Beenakker's theory [4]. Using the data in Fig. 3, one
can quantitatively restore the degree of the current spin-polarization P$_{C}$
for LCMO. We analyzed the results obtained on such type contacts on the basis
of the ballistic and diffusion models [17,25,26]. The restored magnitude of
P$_{C}$ was 75-85\%. (Details will be presented elsewhere, Ref. 24).

For some of cases the measured contact's spectrum reveals very distinct
features which we interpret as the manifestation of an unconventional
proximity effect. The finger-print of such junctions is a quite visible drop
of the contact's resistivity just after superconducting transition of Pb.
Pronounced picture of proximity effect we observed on a few junctions. Figure
2 (main panel) shows an example of temperature dependence of the resistance of
the proximity affected contact. At T
%TCIMACRO{\TEXTsymbol{<} }%
%BeginExpansion
$<$
%EndExpansion
7.2K a sharp drop of the contact's resistivity is observed. Reduction of the
resistance $\delta$R (about 15\%) is by two orders of magnitude larger than it
might be expected from the conventional theory of proximity effect for S/F
contacts. In Fig. 4 both the current-voltage and the conductance G(V) =
($d$I/$d$V)/(I/V) versus voltage dependences for the Pb/LCMO proximity
affected point contact at 4.2K are presented. It is evident that, like for a
conventional AR at S/N interface and opposite to the results in Fig. 3, the
excess current and doubling of the normal-state conductivity has been observed.

For S/N structure, the AR acts as a parallel conduction channel to the initial
electron current, doubling the normal-state conductance G$_{N}$ of the point
contact for applied voltage eV
%TCIMACRO{\TEXTsymbol{<} }%
%BeginExpansion
$<$
%EndExpansion
$\Delta_{1}$, where $\Delta_{1}$ is the single-particle gap at the interface.
As it follows from the experimental results in Fig. 4, the proximity induced
single-particle gap at Pb/LCMO interface, $\Delta_{1}$ $\approx$ 18 meV, is
much larger than that of Pb: $\Delta_{Pb}$(T = 0) = 1.41meV. In Fig. 5 the
temperature dependence of the AR spectra for another proximity affected
Pb/LaCaMnO contact is shown. As it is seen, the supercurrent conversion into
quasiparticle current by AR mechanism takes place up to 7.04K, while the ratio
G(0)/G(eV%
%TCIMACRO{\TEXTsymbol{>}}%
%BeginExpansion
$>$%
%EndExpansion
$\Delta$) is notably decreased. This proves that the superconductivity of LCMO
is due to superconducting state of Pb. Fondly comparing the value of proximity
induced single-particle gap $\Delta_{1}$ $\approx$ 18 meV and T$_{C}$ (= 7.2K
for bulk Pb), one can find that for proximity induced superconducting state of
S/LCMO interface the conventional BCS ratio $\Delta_{BCS}$(T = 0) =
1.76T$_{C}$ does not hold.

We should note that for some contacts we detected the AR spectrum that may be
considered as evidence of coexistence at S/LCMO interface of both conventional
and unconventional superconducting pairing. Representative data for such
contacts is shown in Fig. 6. Also, in Figs. 4 and 5 for voltage eV
%TCIMACRO{\TEXTsymbol{>} }%
%BeginExpansion
$>$
%EndExpansion
$\Delta_{1}$, one can distinguish the peaks which we attributed are due to a
formation of the so-called \textquotedblleft phase-slip
lines\textquotedblright\ -- the two-dimensional analogue of phase-slip centers
[27,28]. Directly the existence of phase slipping is visible in Fig. 7, where
representative excess current-voltage (I$_{exc}$-V) dependence is shown. There
we observe several features which are characteristic to phase-slip lines
[27,28]: (i) a voltage jump at some critical current; (ii) all resistive
branches have the same excess current, given by the intersection of their
slopes with current axis. The value of the excess current is found to fulfill
the theoretical relation I$_{exc}$ = (4/3)$\Delta_{1}$/eR$_{N}$ [29] for
$\Delta_{1}$ $\approx$ 18 meV and experimental value of the contact's normal
state resistance R$_{N}$ = 232 $\Omega$. I.e., the excess current can not be
attributed to Pb electrode. The step-like increasing of the contact's
resistance is due to a discrete destruction (phase-slip) of superconducting
state of surface region LCMO. One can make a crude estimation about the
thickness of this surface region, L$_{SC}$, being in superconducting state.
Indeed, the thickness of the normal-state layer for each phase-slip line is of
the order 2$\xi_{LCMO}$, where $\xi_{LCMO}$ is the superconducting coherence
length of LCMO, $\xi_{LCMO}$ $\approx$ $\hbar$v$_{F}$/$\Delta_{LCMO}$. For
manganites the Fermi velocity v$_{F}$ is $\sim$10$^{7}$sm/sec and taking
naively (see Discussion) $\Delta_{LCMO}\sim$ $\Delta_{1}$ one can easily
obtain $\xi_{LCMO}$ $\approx$ 30\AA . In Fig. 7, six jumps are definitely
detected. I.e., L$_{SC}$ is about 6(2$\xi$LCMO) $\sim$ 300-400\AA . If we
assume that $\Delta_{LCMO}$ is about $\Delta_{Pb}$, then the thickness of the
surface region with proximity induced superconductivity is larger by order.

\textit{Discussions}. -- The exact processes of conversion by which the
supercurrent is continued as the quasiparticle current in an adjacent
half-metal ferromagnetic layer are not known yet. Recent theories proposed a
few mechanisms which may cause a nonvanishing Josephson and long-range
proximity effects in S/HMF/S structures. The most elaborated one is based on
the induced triplet correlations which result in the indirect proximity
effect. In the model [30,31] the presence of the superconducting triplet
correlations with an unusually long penetration length in the ferromagnetic
metal requires the interplay of two separate interface processes: spin mixing
and spin-flip scattering. The spin-rotation effect alone generates at the
superconducting side of the S/HMF boundary the triplet correlation with
\textquotedblleft zero spin\textquotedblright\ component of the form
f$_{\uparrow\downarrow}$ + f$_{\downarrow\uparrow}$ (here f$_{\uparrow
\downarrow}$ and f$_{\downarrow\uparrow}$ stand for opposite-spin triplet pair
amplitudes). Similar to a wave function of the singlet (Cooper) pair, this
component penetrates into HMF on a short distance $\xi_{F}$ $\sim$ ($\hbar
$D$_{F}$/$\pi$H$_{exc}$)$^{1/2}$ $\sim$ ( $\hbar$D$_{F}$/$\pi$\textit{k}$_{B}%
$T$_{Curie}$)$^{1/2}$
%TCIMACRO{\TEXTsymbol{<}}%
%BeginExpansion
$<$%
%EndExpansion%
%TCIMACRO{\TEXTsymbol{<} }%
%BeginExpansion
$<$
%EndExpansion
$\xi_{S}$ (here H$_{exc}$ denotes the exchange energy). Spin-flip scattering
induces both \textquotedblleft nonzero spin\textquotedblright\ triplet
components of the pair amplitudes, f$_{\uparrow\uparrow}$ and f$_{\downarrow
\downarrow}$, in the superconductor, and the f$_{\uparrow\uparrow}$ (or
f$_{\downarrow\downarrow}$) pair amplitude in the half metal ferromagnet. The
equal-spin triplet correlations f$_{\uparrow\uparrow}$ and f$_{\downarrow
\downarrow}$ posses an unusually long penetration length in the ferromagnet,
such as for normal nonmagnetic metal: $\xi$ $\sim$ ($\hbar$D$_{F}$/$\pi
$\textit{k}$_{B}$T)$^{1/2}\sim\xi_{N}$. The long-range triplet correlations
may also arise in S/F structures with a nonuniform magnetization in the
ferromagnet [32-34]. Remarkably, the local magnetic nonhomogeneity of S/F
interface, no matter whether it has been introduced as a \textquotedblleft
spin-active\textquotedblright\ interface or as a nonuniform magnetization, is
a key factor of the models [30-34]. At present, however, no existing
technology can create such nonhomogeneity in a controllable way with nanoscale precision.

On the other hand, several theoretical models and numerous experimental data
suggest that nano-scale nonhomogeneity, referred to as phase separation, is an
intrinsic feature of colossal magnetoresistive manganites (see for example
Ref. [21] and references therein). In particular, the following draft picture
for a surface (thickness of a few nanometers) magnetic structure of manganites
has emerged at present: since the double exchange mechanism is sensitive to a
Mn-O-Mn bound state, any structural disorder (oxygen non-stoichiometry,
vacancies, stress, etc.) near surface region of grain suppresses the double
exchange and leads to a local spin disorder. In particular, for La$_{1-x}%
$Ca$_{x}$MnO$_{3}$ with x $\approx$ 0.3, scanning tunneling spectroscopy [35]
and neutron diffuse scattering [36] indicate the existence of magnetic
nonhomogeneities attributed to hole-rich and hole-poor clusters and even allow
to estimate the shape and the size (a few lattice spacings) as well as the
magnetic structure of the clusters. Another feature important for our
discussion is that, due to strong Hund's interaction (for Mn$^{3+}$ the Hund's
energy $\sim$1eV [21]), spin disorder serves as strong spin-scattering centers
for charge carries.

Incorporating the internal magnetic nonhomogeneity of LCMO, the AR spectra
observations can be explained as follows. For the Sharvin contacts
\textquotedblleft without proximity effect\textquotedblright, the large but
not full spin polarization of transport in LCMO most naturally can be
explained in the model of magnetic phase segregation of the crystal on a
nanometer scale where only one phase corresponds to the state of the
ferromagnetic metal with full charge carriers spin polarization. (The details
of such physics behind the observed incomplete spin polarization of transport
in manganites are discussed in Ref. [24].) For \textquotedblleft proximity
affected contacts\textquotedblright, the observed large change in the
contact's resistance together with its sing can not be explained on the basis
of conventional, i.e., singlet pairing, superconducting proximity effect. We
supposed that in such contacts the conditions for unconventional proximity
effect are fulfilled. I.e., depending on the exact magnetic nonhomogeneity at
S/HMF boundary, the LCMO surface causes superconducting triplet f$_{\uparrow
\uparrow}$ (or f$_{\downarrow\downarrow}$) correlations which decay slowly
into the half-metal. Being proximity induced, the supercurrent of equal-spin
triplet pairs is continued as a quasiparticle current in a bulk of the
half-metal ferromagnetic layer through a usual Andreev reflection mechanism,
with an excess current and doubling of the normal-state conductance for
applied voltage less than the single-particle gap.

At this stage we can only speculate about the discrepancy between the
proximity induced single-particle gap at Pb/LCMO interface and the
superconducting gap of Pb. The value of the single-particle gap $\Delta_{1}$
obtained from the AR spectra may be qualitatively justified by generating the
models of superconductivity of strongly disordered (granular) superconductors
[37,38] on our case of proximity induced triplet superconductivity of
phase-separated manganites. It has been known for many years (see, e.g., [39])
that some superconductors, when composed of small grains, show the $\Delta
$/T$_{C}$ ratio higher than the BCS value. Intuitively, it is clear that the
proximity induced superconducting state of a manganite with hole-rich and
hole-poor clusters of a few lattice spacing is practically the same as that of
a strongly disordered metal (or metal nanoparticles) with attractive
interaction and superconductivity due to a tender coherence between localized
Cooper pairs. The strength of disorder can be characterized by two relevant
energy scales: $\delta$ and $\Delta_{1}$, where $\delta$ is the effective
level spacing in the localization volume, and $\Delta_{1}$ is the
single-particle gap. In the regime $\delta$
%TCIMACRO{\TEXTsymbol{<}}%
%BeginExpansion
$<$%
%EndExpansion%
%TCIMACRO{\TEXTsymbol{<} }%
%BeginExpansion
$<$
%EndExpansion
$\Delta_{Pb}$
%TCIMACRO{\TEXTsymbol{<}}%
%BeginExpansion
$<$%
%EndExpansion%
%TCIMACRO{\TEXTsymbol{<} }%
%BeginExpansion
$<$
%EndExpansion
$\Delta_{1}$ superconductivity persists due to induced triplet pairs. However,
due to an extra cost of creating an unpaired electron, the energy required to
add an electron in this regime is no longer equal to self-consistent BCS gap
and the ratio of this gap to T$_{C}$ can be anomalously large [38]. I.e, in
such system, the superconducting pairing energy does not characterize the
spectral gap and the typical single-particle gap $\Delta_{1}$, that most
naturally is measured by tunnelling or AR spectroscopy experiments, is much
larger that T$_{C}$.

\textit{Conclusion}. -- In summary, to study superconducting pairing in
heterostructures of low-temperature superconductor and a half-metallic
ferromagnet, we have prepared the point contacts of Pb/La$_{0.65}$Ca$_{0.35}%
$MnO$_{3}$ where the Andreev reflection plays an important role. Several
evidence of the existence of unconventional superconducting pairing and
long-range proximity effect are detected. In particular, since the
supercurrent of equal-spin pairs can be continued as quasiparticle current in
a bulk of a half-metal ferromagnet due to usual AR mechanism, such
finger-print of AR as excess current and doubling of the normal-state
conductance have been observed.\ The obtained experimental data can be
understood within a model of inherent magnetic nonhomogeneity of manganites if
one includes concepts (i) of proximity induced superconducting triplet
correlations at S/HMF interface with a long-range decay length, and (ii)
superconductivity of a phase separated (hole-rich and hole-pore clusters)
surface of manganite due to a fragile coherence between localized triplet
pairs. It appears that under-doped manganites as well as region at surface of
metallic ferromagnetic manganites are convenient materials to reveal new
mechanisms of superconductivity. The results obtained are of great relevance
for spin-electronics devices based on exploration of the nanoscale magnetism
of manganites and their half-metallic features.

Authors acknowledge the participants of the ICFM-05 (Partenit, Crimea,
Ukraine) for valuable discussion.

\begin{center}
--------------------------------------------------------
\end{center}

1. A. A. Golubov, M. Y. Kupriyanov, and E. Il'ichev, Rev. Mod. Phys.
\textbf{76}, 411 (2004).

2. A. Buzdin, Rev. Mod. Phys. \textbf{77}, 935 (2005).

3. F. S. Bergeret, A. F. Volkov, and K. B. Efetov, Rev. Mod. Phys.
\textbf{77}, \#4 (2005).

4. M. J. M. de Jong and C. W. J. Beenakker, Phys. Rev. Lett. \textbf{74}, 1657 (1995).

5. M. Kasai, Y. Kanke, T. Ohno, and Y.Kozono, J. Appl. Phys. \textbf{72}, 5344 (1992).

6. V. T. Petrashov, N. V. Antonov, S. Maksimov, and R. Shaikhaidarov, JETP
Lett. \textbf{59}, 551(1994).

7. V. T. Petrashov, I. A. Sosnin, I. Cox, A. Parsons, and C. Troadec. Phys.
Rev. Lett. \textbf{83}, 3281 (1999).

8. Z. Sefrioui, D. Arias, V. Pe\v{n}a, J. E. Villagas, M. Varela, P. Prieto,
C. Leon, J. L. Martinez, and J. Santamaria, Phys. Rev. B \textbf{67}, 214511 (2003).

9. V. Pe\v{n}a, Z. Sefrioui, D. Arias, C. Leon, J. Santamaria, M. Varela, S.J.
Pennycook, and J. L. Martinez, Phys. Rev. B \textbf{69}, 224502 (2004).

10. K. Senapati and R. C. Budhani, cond-mat/0507073 (2005).

11. I. Sonin, H. Cho, V. T. Petrashov, and A. F. Volkov, cond-mat/0511077 (2005).

12. E. L. Wolf, Principles of Electron Tunneling Spectroscopy (Oxford
University Press, New York, 1985).

13. A. F. Andreev, Sov. Phys. JETP, \textbf{19}, 1228 (1964).

14. J. M. Byers and M. E. Flatte, Phys. Rev. Lett. \textbf{74}, 306 (1995).

15. D. Beckmann and H. B. Weber, Phys. Rev. Lett. \textbf{93}, 197003 (2004).

16. R. J. Soulen Jr., J. M. Byers, M. S. Osofsky, B. Nadgorny, T. Ambrose, S.
F. Cheng, P. R. Broussard, C. T. Tanaka, J. Nowak, J. S. Moodera, A. Barry,
and J. M. D. Coey, Science \textbf{282}, 85 (1998).

17. B. Nadgorny, I. I. Mazin, M. Osofsky, R. J. Soulen, Jr., P. Broussard, R.
M. Stroud, D. J. Singh, V. G. Harris, A. Arsenov, and Ya. Mukovskii, Phys.
Rev. B \textbf{63}, 184433 (2001).

18. Y. Ji, G. J. Strijkers, F. Y. Yang, C. L. Chien, J. M. Byers, A.
Anguelouch, G. Xiao, and A. Gupta, Phys. Rev. Lett. \textbf{86}, 5585 (2001).

19. G. J. Strijkers, Y. Ji, F. Y. Yang, C. L. Chien, and J. M. Byers, Phys.
Rev. B \textbf{63}, 104510 (2001).

20. A. Hoffmann, S. G. E. Te Velthuis, Z. Sefrioui, J. Santamaria, M. R.
Fitzsimmons, S. Park, and M. Varela, Phys. Rev. B \textbf{72}, 140407 (2005).

21. E. Dagotto, T. Hotta, and A. Moreo, Phys. Rep. \textbf{344}, 1 (2001).

22. G. Wexler, Proc. Phys. Soc. London \textbf{89}, 927 (1966).

23. Yu. V. Sharvin, JETP \textbf{21}, 655 (1965).

24. A.I. D'yachenko, V.A. D'yachenko, V. Yu. Tarenkov, and V. N. Krivoruchko,
Fiz. Tverd. Tela \textbf{48}, \#3 (2006).

25. I. I. Mazin, A. A. Golubov, and B. Nadgorny. J. Appl. Phys. \textbf{89},
7576 (2001).

26. B. P. Vodopyanov, and L. R. Tagirov, JETP Lett. \textbf{77}, 126 (2003).

27. I. M. Dmitrenko, Low. Temp. Phys. \textbf{22}, 686 (2001).

28. A. G. Sivakov, A. M. Glukhov, A. N. Omelyanchouk, Y. Koval, P. M\"{u}ller,
and A. V. Ustinov, Phys. Rev. Lett. \textbf{91}, 267001 (2003).

29. G. E. Blonder, M. Tinkham, and T. M. Klapwijk, Phys. Rev. B \textbf{25},
4515 (1982).

30. M. Eschring, J. Kopu, J. C. Cuevas, and G. Sch\"{o}n, Phys. Rev. Lett.
\textbf{90}, 137003 (2003).

31. T. Champel and M. Eschrig, B \textbf{72}, 064523 (2005).

32. F. S. Bergeret, A.F. Volkov, and K. B. Efetov, Phys. Rev. Lett.
\textbf{86}, 4096 (2001); Phys. Rev. B \textbf{64}, 134506 (2001).

33. A. Kadigrobov, R. I. Skekhter, and M. Jonson, Europhys. Lett. \textbf{54},
394 (2001).

34. A. F. Volkov, Ya. V. Fominov, and K. B. Efetov, Phys. Rev. B \textbf{72},
184504 (2005).

35. M. F\"{a}th, S. Freisem, A. A. Mewnovsky, Y. Tomioka, J. Aarts,\ and J. A.
Mydosh, Science \textbf{285}, 1540 (1999).

36. M. Hennion, F. Moussa, P. Lehouelleur, F. Wang, A. Ivanov, Y. M.
Mukovskii, and D. Shulyatev, Phys. Rev. Lett. \textbf{94}, 057006 (2005).

37. V. F. Gantmakher, Physics-Uspekhi \textbf{41}, 214 (1998).

38. M. Feigel'man, L. B. Ioffe and E. A. Yuzbashyan, cond-mat/0504766 (2005).

39. R. W. Cohen and B. Abeles, Phys. Rev. \textbf{168}, 444 (1968).

\begin{center}

Figure Captures
\end{center}

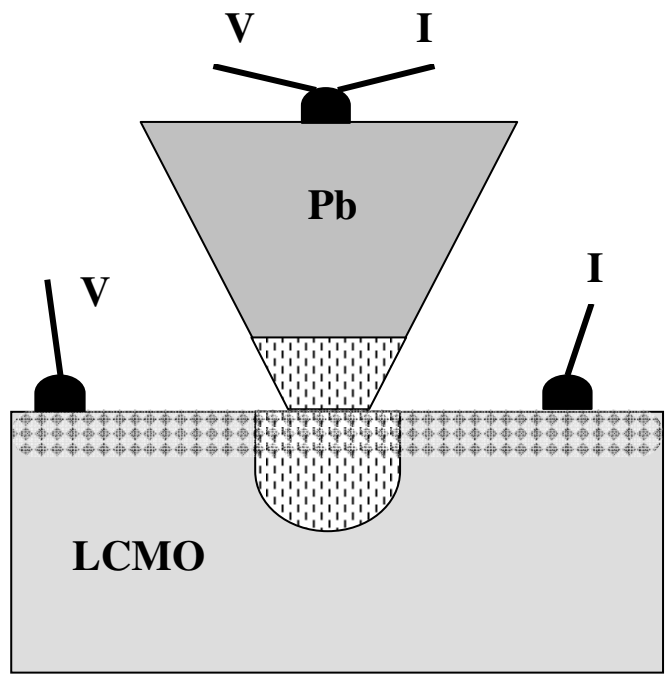
FIG. 1. Geometry of point contacts under consideration. Proximity affected
regions, and magnetic inhomogenuity (hole-rich and hole-pore clusters) of the
manganite surface are shown schematically.

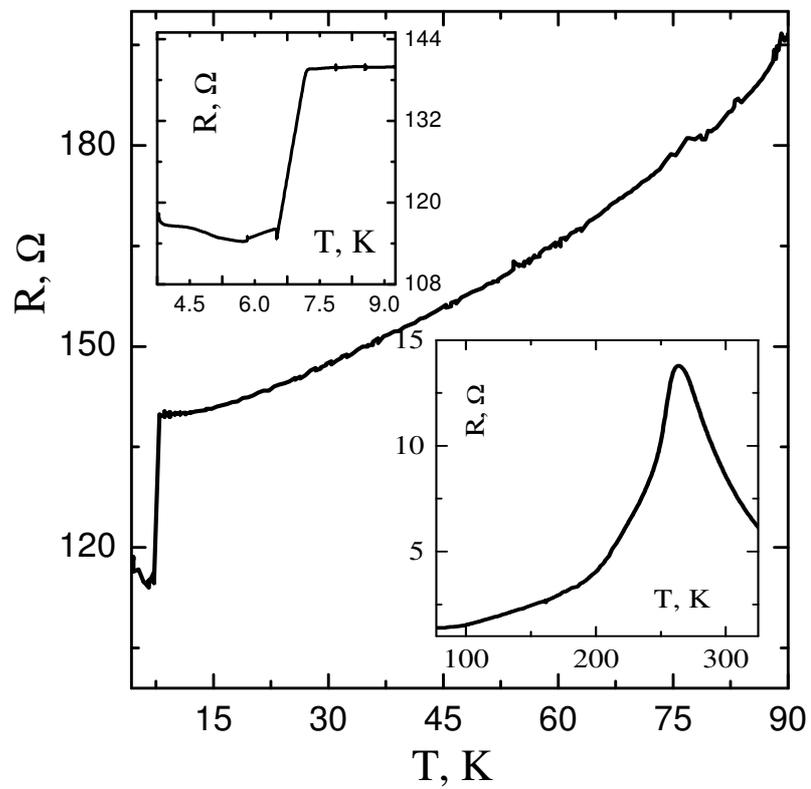
FIG. 2. Temperature dependence of the resistance of the proximity affected
contact. Upper insert: temperature dependence of the resistance of the
proximity affected contact in the region of Lead superconducting transition.
Bottom insert: temperature dependence of the LCMO plates' resistance.

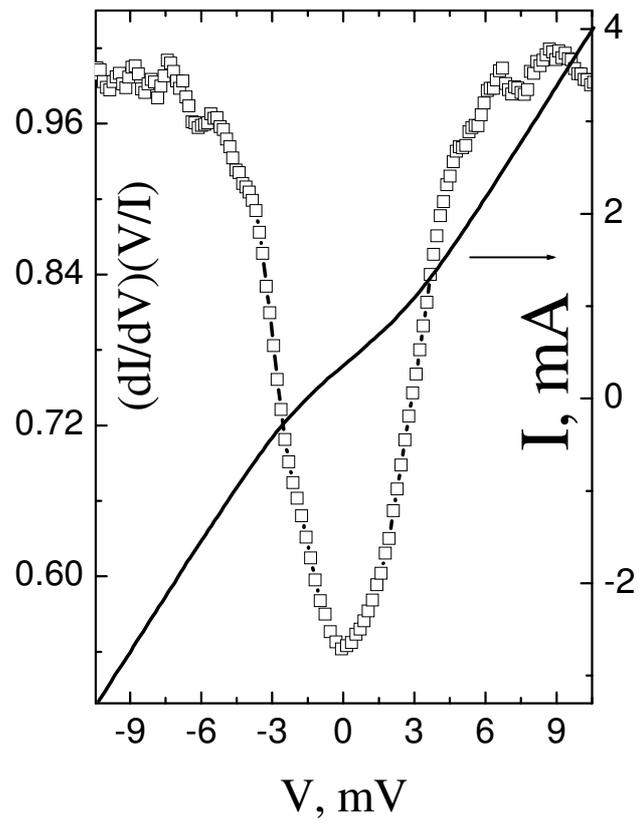
FIG. 3. Current-voltage and normalized differential conductance G(V) =
($d$I/$d$V)(V/I) of the Pb/LCMO contact without visible superconducting
proximity effect; T = 4.2 K.

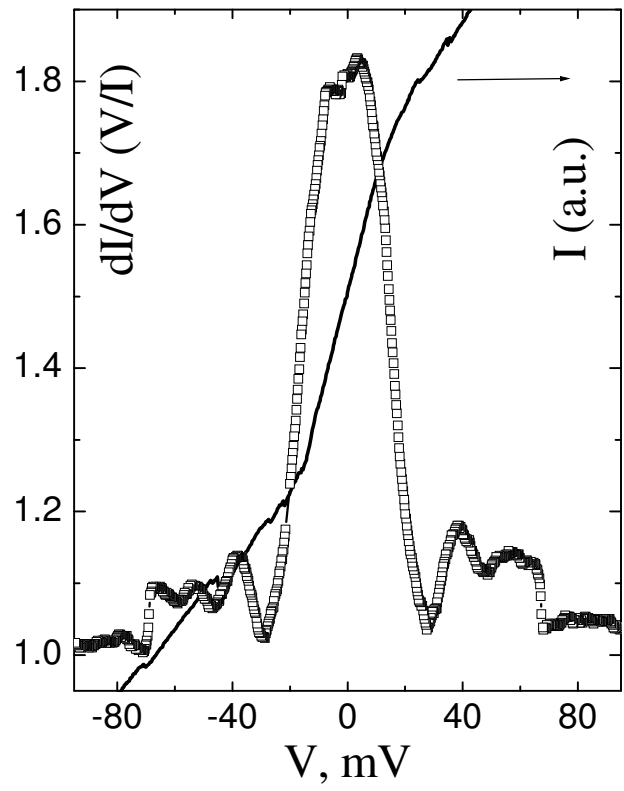
FIG. 4. Current-voltage and normalized conductance G(V) = ($d$I/$d$V)/(I/V) vs
voltage dependences for proximity affected Pb/LCMO point contact at 4.2K

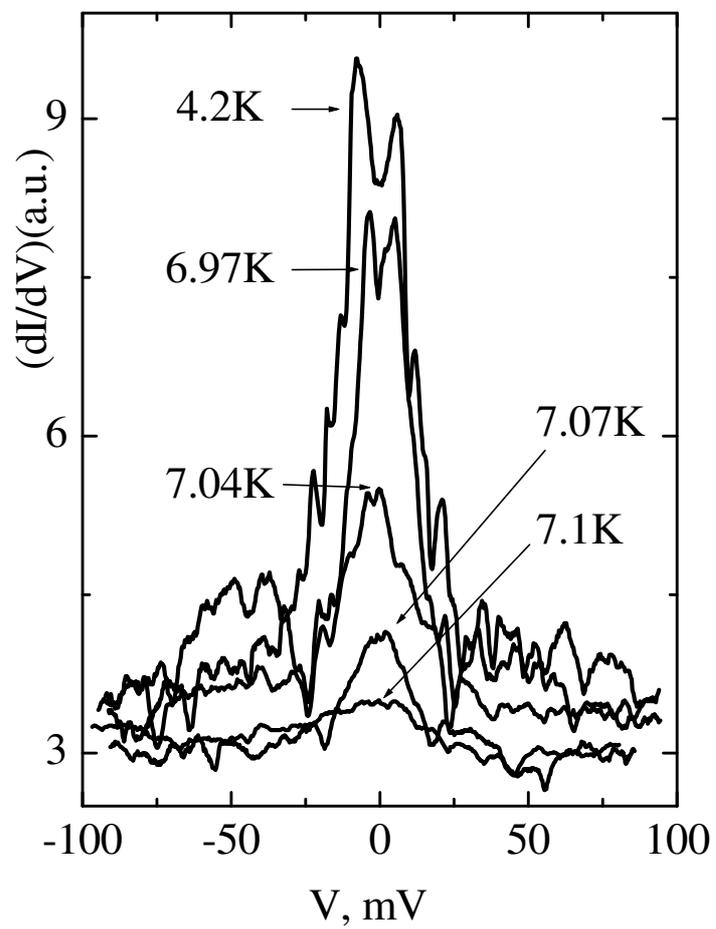
FIG. 5. Temperature dependence of the AR spectra for proximity affected
Pb/LaCaMnO contact. The curves are shifted for clearness.

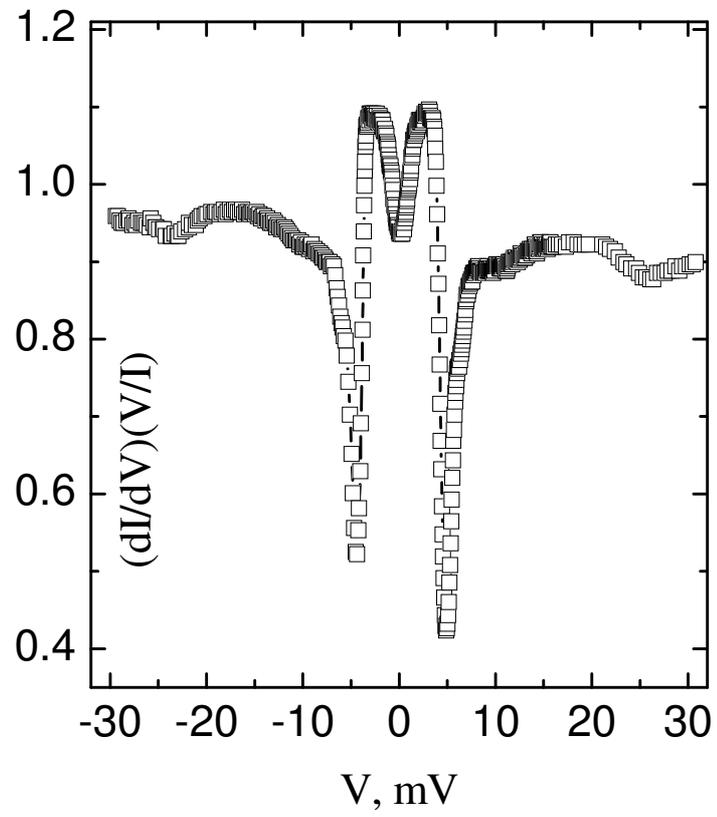
FIG. 6. An example of complex AR spectra has been observed for some Pb/LCMO contacts.

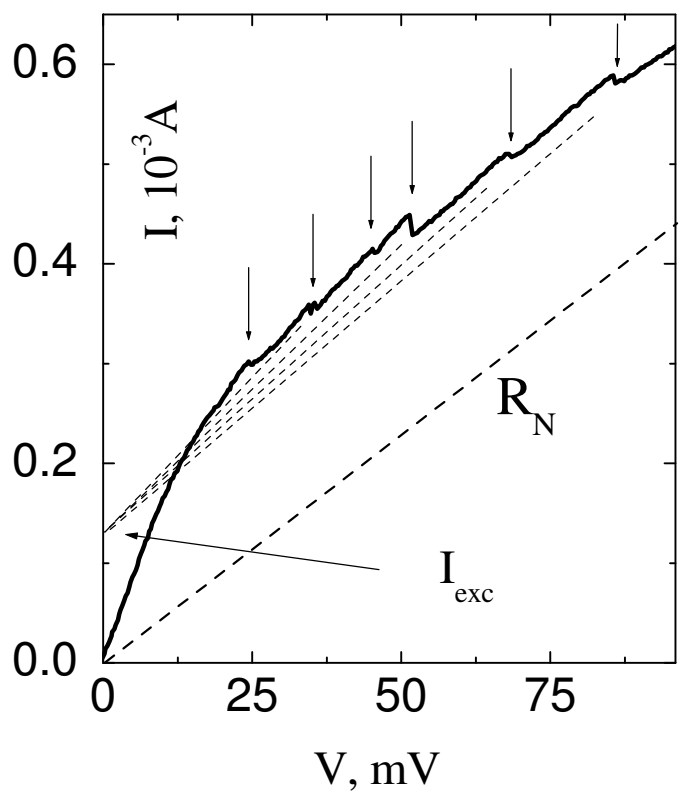
FIG. 7. The excess current-voltage (I$_{exc}$-V) characteristic for proximity
affected contact. Each voltage jump in I$_{exc}$-V curve correspods to a
generation of an additional phase-slip line in the surface region of LCMO.

\end{document}